\documentclass[aps,prc,twocolumn,showpacs,superscriptaddress,groupedaddress,a4paper]{revtex4}
\usepackage{graphicx}
\usepackage{epstopdf,color,bm}
\begin{document}
\title{Coulomb-nuclear dynamics in the weakly-bound $^8$Li breakup}
\author{B. Mukeru}
\affiliation{Department of Physics, University of South Africa, P.O. Box 392, Pretoria 0003, South Africa}
\author{J. Lubian}
\affiliation{Instituto de F\'isica, Universidade Federal Fluminense, 
Niter\'oi, RJ, 24210-340, Brazil}
\author{Lauro Tomio}
\affiliation{Instituto de F\'isica T\'eorica, Universidade Estadual Paulista, 01140-070 S\~ao Paulo, SP, Brazil}

\begin{abstract}
A detailed study of total, Coulomb and nuclear breakup cross sections dependence on the projectile 
ground-state binding energy $\varepsilon_b$ is presented, by considering the $^8$Li+$^{12}$C and 
$^8$Li+$^{208}$Pb breakup reactions.  To this end,  apart from the experimental one-neutron separation 
energy of $^8$Li nucleus ($\varepsilon_b=2.03$~MeV), lower values of $\varepsilon_b$ down to 
$\varepsilon_b=0.01$~MeV, are also being considered.  It is shown that all breakup processes become 
peripheral as $\varepsilon_b\to 0.01$ MeV, which is understood as due to the  well-known large spacial 
extension of ground-state wave functions associated to weakly-bound projectiles. 
The Coulomb breakup cross section is found to be more strongly dependent on $\varepsilon_b$ than  
the nuclear breakup cross section, such that the Coulomb breakup becomes more significant as 
$\varepsilon_b$ decreases, even in a naturally nuclear-dominated reaction. This is mainly due to the 
long-range nature of the Coulomb forces, leading to a direct dependence of the Coulomb breakup on the 
electromagnetic transition matrix. It is also highlighted the fact that the nuclear absorption plays a minor 
role for small binding when the breakup becomes more peripheral.\\
\end{abstract}
\maketitle

\section{Introduction}
In the breakup of weakly-bound projectiles against heavy nuclei targets, a relevant phenomenon which has been investigated 
is the Coulomb-nuclear dynamics, such that considerable efforts have been made to understand the role of Coulomb-nuclear 
interference and the dynamics of fragments absorption in the breakup process. The established studies in this matter 
with the corresponding most relevant works can be found in  Refs.~\cite{2003Suzuki,Thompson100,Chat10}.
For other complementary studies done in past two decades on the Coulomb-nuclear dynamics involving weakly-bound 
projectiles, with particular interest to our present investigation, we select the  
Refs.~\cite{Thomp20,1999Th,2002Margueron,2003Capel,Tarutina10,Hussein20,2006Canto,2009Lubian,2009Canto},
as well as more recent works (among which we include contributions by some of us) in 
Refs.~\cite{Kucuk10,2014Capel,Hussein200,2015Lubian,Mukeru10,Mukeru20,2016Manjeet,Pierre100,2017Mukeru,
MukeruPRC2020}.
Despite the advances verified by these studies, the question on how both Coulomb and nuclear forces interfere to produce 
a total breakup remains far from being fully established.  Some of the challenges emanate from the fact that, 
in a Coulomb-dominated reaction, small contribution of the nuclear breakup does not automatically imply insignificant 
Coulomb-nuclear interference~\cite{Nakamura10,Noc10,Aum10,Fukuda10,Abu10}. 
It could be interesting to verify what happens in nuclear-dominated reactions.

In view of the long-range nature of Coulomb forces, a low breakup threshold is expected to lead to peripheral collisions, 
where the Coulomb breakup prevails over the nuclear breakup.  In this peripheral region, the Coulomb breakup cross section
depends on the projectile structure through the electromagnetic matrix elements of the projectile.  Although 
not a general rule,  according to the 
Coulomb dissociation method~\cite{Bertulani10,Winther10,Baur10,Baur20}, the breakup cross section is simply the product 
of the reaction parameters and the projectile dipole electric transition probability.  As the binding energy decreases, the 
reaction becomes more peripheral, with the ratio between the Coulomb breakup cross section to the nuclear counterpart 
being expected to rise significantly, regardless the target mass. Intuitively, in this case,  one would expect that the total breakup 
cross section becomes comparable to the Coulomb one, owing to both dynamic and static breakup effects. From the fact that
lower is the ground-state binding, longer is the tail of the associated wave function, the nuclear forces are fairly stretched 
beyond the projectile-target radius. Therefore, for a projectile with very weak binding energy, even the nuclear breakup can be 
assumed to be a peripheral phenomenon, with the Coulomb-nuclear interference becoming stronger in the peripheral region.

The dependence of various reaction observables on the projectile ground-state binding energy has being studied
recently in  Refs.~\cite{Wang10,Rath100,2016Rangel,Lei100,Mukeru15,2020Mukeru},  in which different projectiles with 
different binding energies have been considered.  One of the drawbacks being that all the projectiles do not have the same 
 ground-state structure, mass and charge.  Among other ways to 
circumvent such shortcomings, at least theoretically, one could artificially consider different binding energies for the same 
projectile (i.e., with nucleon-number $A$ and charge $Z$ unchanged), within an approach that has been adopted for instance 
in Refs.~\cite{2016Rangel,Lei100,Mukeru15,2020Mukeru}.  Even though a given nucleus has fixed ground-state energy,  this is 
a convenient theoretical approach to unambiguously establish the dependence of the reaction observables on the projectile 
binding energy.

Another important aspect in breakup dynamics, relies on possible effects on other reaction observables, such as on 
fusion cross sections. While it is widely understood that the complete fusion suppression strongly depends on the projectile breakup threshold (see Ref.~\cite{2020Jha}, for recent related studies), 
strong charge clustering has recently been identified as the main factor responsible for such suppression, in the breakup of $^8$Li on a heavy target \cite{Cook20}. Some 
 behavior 
in the breakup of this nucleus, has also been  reported in Refs.~\cite{Pak20,Gum20}. Unlike several other loosely bound nuclei 
(such as $^8$B, $^{6,7}$Li, and $^{11}$Be), not much has been reported on the breakup dynamics of the 
$^8{\rm Li}$ nucleus.

In view of the above discussion, we are motivated to study the breakup of $^8{\rm Li}$ nucleus, within a model in which a 
valence neutron (n) is loosely bound to the $^7{\rm Li}$ nucleus by a binding energy $\varepsilon_b=2.03$ MeV~\cite{Nut10}, 
by considering the light and heavy targets $^{12}$C and $^{208}$Pb.
The present study on the $^8$Li$+^{208}$Pb breakup reaction is also extending a previous recent analysis
for this reaction done in Ref.~\cite{2020Mukeru}, where a critical angular momentum for complete fusion was also 
 considered.
We are particularly interested in analyzing  the dependence of the resulting total, Coulomb and nuclear breakup cross 
 sections,  as well as the Coulomb-nuclear interference, on the projectile ground-state binding energy, in order to test 
 the validity of the 
assumptions presented in the previous paragraphs.
Within a more detailed investigation, we expect to show that for a much weaker projectile binding energy, the Coulomb 
breakup becomes dominant regardless the target mass, and the nuclear breakup becomes relatively peripheral, leading to a 
peripheral Coulomb-nuclear interference. Since both Coulomb and nuclear breakup cross sections increase with the decrease 
of the binding energy, a clear separation of their effects is not a simple task. The choice of $^{12}$C and $^{208}$Pb as the 
targets is motivated by the fact that, in the former case, the reaction should be dominated by the nuclear breakup, whereas it 
is dominated by Coulomb breakup in the latter case.  If fact, $^{12}$C was also used in Ref.~\cite{Fukuda10}, as a reference 
target when studying the $^{11}$Be Coulomb dissociation on $^{208}$Pb target. In our approach to obtain the 
corresponding total, Coulomb and nuclear breakup cross sections, we adopt the Continuum Discretized Coupled 
Channels (CDCC) formalism~\cite{Aust100}, with the Fresco code~\cite{1988Thompson} being used for the 
numerical solutions.
 
The next sections are organized as follows: Sect.~\ref{calculation} provides some details on the model approach, with a 
summary of the CDCC formalism. Sect.~\ref{results} contains the main results for elastic and breakup cross sections, 
 together with our analysis on the Coulomb-nuclear interference and possible absorption contributions. Finally, the 
Sect.~\ref{conclusion} presents a summary with our conclusions.

\section{Formalism and computational approach}
\label{calculation}
\subsection{Brief description of the CDCC formalism}
As mentioned in the introduction, in our numerical approach we use the CDCC formalism, in which we model
the projectile $^8{\rm Li}$ as $^7{\rm Li}$ core nucleus, to which a neutron is loosely bound 
with ground-state energy 
$\varepsilon_b=2.03\,{\rm MeV}$.  This state is defined in the core-neutron centre-of-mass (c.m.) by 
$n=1$, $\ell_0=$1, $\tilde{\j}_0^{\pi}=2^+$ quantum numbers, where $n$ stands for the radial state, 
$\ell_0$ the orbital 
angular momentum and $\tilde{\j}_0^\pi$ the projectile total angular momentum with parity $\pi$. 
It is obtained 
by applying the usual spin-orbit coupling ${\bm j}_0={\bm \ell}_0+{\bf 1/2}$; 
$\tilde{\bm \j}_0={\bm j}_0+{\bm I}_c$, 
with the core spin $I_c={3}/{2}$.
In addition to the ground state, an excited bound state with energy $\varepsilon_{\rm ex}=0.98\,{\rm MeV}$ 
(located in the $\tilde{\j}_0^{\pi}=1^+$ state~\cite{Nut10})  was also considered in our coupling scheme.
We would like to emphasize that we are not considering possible core excitations in our calculations.

In this formalism, we first consider the expansion of the three-body wave function
on the projectile internal states. After that, by introducing the three-body expansion 
into the corresponding Schr\"odinger equation, a one-dimensional radial set of coupled 
differential equations can be derived for the radial wave-function components 
$\chi_{\alpha}^{LJ}(R)$, in terms of the projectile-target c.m. coordinate
$R$, which is given by
\begin{eqnarray}\label{coupled}
&&\left[-\frac{\hbar^2}{2\mu_{pt}}\bigg(\frac{d^2}{dR^2}-\frac{L(L+1)}{R^2}\bigg)+
 U_{\alpha\alpha}^{LLJ}(R)\right]
\chi_{\alpha}^{LJ}(R)\nonumber\\
&&+\sum_{\alpha'L' (\alpha'\ne\alpha)}
U_{\alpha\alpha'}^{LL'J}(R)\chi_{\alpha'}^{L'J}=(E-\varepsilon_\alpha)\chi_{\alpha}^{LJ},
\end{eqnarray}
where $L$ is the orbital angular momentum associated with $R$, $J$ is the total angular momentum, and
$\mu_{pt}$ the projectile-target ($pt$) reduced mass. The total energy is given by $E$, with 
$\varepsilon_{\alpha}$ being the projectile bin energies. The index $\alpha$ appearing in the equation is representing 
a set of quantum numbers describing the projectile states, as given by 
$\alpha\equiv (i,\ell,s,j,I_c,\tilde{\j})$,  $i=0,1,2,\ldots,N_b$ ($N_b=$ number of bins). 

With the projectile-target potential given as a sum of the core-target ($ct$) and 
neutron-target ($nt$) terms, i.e.,
$U_{pt}({\bm r}, {\bm R})=U_{ct}({\bm R}_{ct})+U_{nt}({\bm R}_{nt})$, where $ {\bm R}_{ct}
\equiv{\bm R}+\frac{1}{8}{\bm r}$ 
and ${\bm R}_{nt}\equiv{\bm R}-\frac{7}{8}{\bm r}$ (with ${\bm r}$ being the projectile internal coordinate),
the potential matrix elements $U_{\alpha\alpha'}^{LL'J}(R)$ in (\ref{coupled}) are given by 
its Coulomb and nuclear parts, such that
{\small \begin{eqnarray}\label{potcomp}
U_{\alpha\alpha'}^{LL'J}(R)
&=&\langle\mathcal{Y}_{\alpha L}({\bm r},\Omega_R)|V_{ct}^{Coul}({\bm R}_{ct})
|\mathcal{Y}_{\alpha' L'}({\bm r},\Omega_R)\rangle\nonumber\\
&+&\langle\mathcal{Y}_{\alpha L}({\bm r},\Omega_R)|U_{ct}^{nucl}({\bm R}_{ct})
|\mathcal{Y}_{\alpha' L'}({\bm r},\Omega_R)\rangle\\
&+&\langle\mathcal{Y}_{\alpha L}({\bm r},\Omega_R)|U_{nt}^{nucl}({\bm R}_{nt})
|\mathcal{Y}_{\alpha' L'}({\bm r},\Omega_R)\rangle\nonumber,
\end{eqnarray}    
}where $\mathcal{Y}_{\alpha L}({\bm r},\Omega_R)\equiv[\hat\Phi_{\alpha}({\bm r})\otimes 
{\rm i}^LY_L^{\Lambda}(\Omega_R)]_{JM}$ is the direct product of the angular part of ${\bm R}$ with the
projectile channel wave function, $\hat\Phi_{\alpha}({\bm r})$, 
which contains the square integrable discretized bin wave functions.
The nuclear terms express the sums of real and imaginary parts.   The former are responsible 
for the nuclear dissociation, whereas the latter accounts for the nuclear absorption. 
 These nuclear terms are, respectively, given by 
      $U_{ct}^{nucl}({\bm R}_{ct})=V_{ct}^{nucl}({\bm R}_{ct})+{\rm i}W_{ct}^{nucl}({\bm R}_{ct})$ and
      $U_{nt}^{nucl}({\bm R}_{nt})=V_{nt}^{nucl}({\bm R}_{nt})$+ ${\rm i}W_{nt}^{nucl}({\bm R}_{nt})$, with  
the Woods-Saxon shape being adopted for both components.
The diagonal coupling matrix elements  $U_{\alpha\alpha}^{LL J}(R)$, contain the monopole nuclear term 
in the projectile-target c.m., which we denote by  
$V_{\beta_0\beta_0}^{LJ}(R)=\langle \Phi_{\beta_0}({\bm r})|U_{ct}^{nucl}+U_{nt}^{nucl}|\Phi_{\beta_0}({\bm r})\rangle$,
where $\beta_0$ represents the set of ground-state projectile quantum numbers, 
$\beta_0\equiv (k_0,\ell_0,s,j_0,I_c,\tilde{\j}_0)$. 
The imaginary part accounts for the absorption in the projectile-target c.m. motion.
    
The separation of the Coulomb and nuclear interactions to obtain the Coulomb and nuclear breakup
cross sections ($\sigma_{Coul}$ and $\sigma_{nucl}$, respectively) remains a challenge in nowadays theories, 
making an accurate description of the Coulomb-nuclear interference a more tricky task. 
For that, in this work we resort to an approximate approach, as follows: 
The nuclear breakup cross sections, defined as $\sigma_{nucl}$, are obtained by including in the coupling matrix elements,
the nuclear components of  $U_{ct}$ and $U_{vt}$ potentials,  plus the diagonal monopole Coulomb potential. On the other hand,  
the Coulomb breakup cross sections, defined as $\sigma_{Coul}$, are 
obtained by including in the matrix elements the Coulomb component of the projectile-target potential, i.e., 
$V_{ct}^{Coul}(R_{ct})$ (as $V_{nt}^{Coul}=0$), plus the monopole nuclear potential.  The total 
breakup cross sections $\sigma_{tot}$ are obtained by including the full $U_{pt}$ potential in the calculations.

Since the early works on Coulomb and nuclear breakup studies~\cite{Thomp20,1999Th},  this approach has been 
widely adopted to study Coulomb and nuclear breakup cross sections, as one can follow from the review~\cite{2015Canto} 
(and references therein). In Ref.~\cite{Pierre100}, where different methods are considered in order to decompose the total 
breakup into its Coulomb and nuclear components, this approach is also referred as {\it weak-coupling approximation}.
Two methods emerged from their discussion, which they refer to as {\it method 1} and {\it method 2}.  
The weak-coupling approximation is very close to {\it method 1} for nuclear breakup,  and close to {\it method 2} 
for Coulomb breakup.
While this approximate procedure will not completely eliminate the ambiguities surrounding the separation of the total 
breakup cross section into its Coulomb and nuclear components (as also outlined in Ref.\cite{Pierre100}),  we believe that it 
is particularly justified in the present work, since by using the $^{12}$C target, the breakup is naturally dominated  
by nuclear dissociation, whereas by using the $^{208}$Pb target the breakup is dominated by Coulomb dissociation.

Once the matrix elements (\ref{potcomp}) are computed, the coupled Eq.~(\ref{coupled}) is solved with the usual 
asymptotic conditions, which for $k_{\alpha}\equiv\sqrt{{(2\mu_{pt}/\hbar^2)(E-\varepsilon_{\alpha})}}$ is given by
\begin{eqnarray}\label{BC}
  \chi_{\alpha}^{LJ}(R)\stackrel{R\to\infty}\longrightarrow \frac{\rm i}{2}\left[H_{\alpha}^-(k_{\alpha}R)\delta_{\alpha\alpha'}-
  H_{\alpha}^+(k_{\alpha}R)S_{\alpha\alpha'}^{LL'J}\right],
\end{eqnarray}
where $H_{\alpha}^{\mp}(k_{\alpha}R)$ are the usual incoming (-) and outgoing (+) Coulomb Hankel 
functions~\cite{Abramo100}, with $S_{\alpha\alpha'}(k_{\alpha})$ being the scattering S-matrix elements.
Due to the short-range nature of nuclear forces, the matrix elements corresponding to the nuclear interaction 
in Eq.~(\ref{potcomp}) will vanish at large distances, ${R\gg R_n}$, where 
\begin{equation} \label{potnucl}
R_n\equiv r_0(A_p^{1/3}+A_t^{1/3}) +\delta_R(\varepsilon_b)\equiv R_0 + \delta_R(\varepsilon_b)
\end{equation}
determines the range of the nuclear forces ($r_0$ being the nucleon size, with $r_0A_{p}^{1/3}$ and 
$r_0A_{t}^{1/3}$ the projectile and target sizes, respectively).
The function $\delta_R(\varepsilon_b)$ is introduced to take into account the 
well-known effect which occurs in weakly-bound systems (low breakup thresholds), as in
halo nuclei, in which the nuclear forces can be stretched beyond $R_0=r_0(A_p^{1/3}+A_t^{1/3})$.  
The various breakup cross sections are obtained by using the relevant S-matrix, as outlined for example 
in Ref.~\cite{Thompson100}.

At large distance ($R\to\infty$),  Eq.(\ref{potcomp}) contains only the Coulomb interaction, which can be expanded 
as~\cite{Hussein50}
 {\small \begin{eqnarray}\label{CE}
  V^{Coul}({\bm r},{\bm R})\stackrel{R\to\infty}\longrightarrow 4\pi Z_te\sum_{\lambda=0}^{\lambda_{\rm max}}
  \frac{\sqrt{2\lambda+1}
  }{R^{\lambda+1}}\left[\mathcal{O}_{\lambda}^{\epsilon}({\bm r})\otimes Y_{\lambda}(\Omega_R)\right]^{0},
  \end{eqnarray}
  }where $Z_te$ is the target charge, with $\lambda$ the multipole order truncated by $\lambda_{\rm max}$.
$\mathcal{O}_{\lambda}^{\epsilon}({\bm r})$ is the projectile electric operator, given by
{\small 
\begin{eqnarray}\label{EO}
\mathcal{O}_{\lambda\mu}^{\epsilon}({\bm r})&=&
\left[Z_c e\left(-\frac{A_n}{A_p}\right)^{\lambda}\right]r^{\lambda}Y_{\lambda}^{\mu}(\Omega_r)=
Z_{\lambda}r^{\lambda}Y_{\lambda}^{\mu}(\Omega_r),
\end{eqnarray}
}where $Z_ce$ is the charge of the projectile core, with $Z_\lambda$ being defined as the effective charge.
The projectile electric transition probability for the transition from the projectile ground-state to the continuum 
states can be obtained through $\mathcal{O}_{\lambda}^{\epsilon}({\bm r})$~\cite{Bertulani50}. For 
excitation energies $\varepsilon$, the  corresponding variation of the electric transition probability 
$B(E\lambda)$ can be written as
{\small
  \begin{eqnarray}\label{elec}
 \frac{dB(E\lambda)}{d\varepsilon}&=&\frac{\mu_{cn}}{\hbar^2 k}\sum_{\tilde{\j}}(2{\tilde{\j}}+1)
 \left|\langle \Phi_{\beta_0}({\bm r})|\mathcal{O}_{\lambda}^{\epsilon}({\bm r})|\Phi_{\beta}({\bm r})\rangle\right|^2,
\end{eqnarray}
}where ($\beta$) refers to the set of quantum numbers in the continuum states
 $\beta\equiv (k,\ell,s,j,I_c,\tilde{\j})$], 
$k=\sqrt{2\mu_{cn}\varepsilon/\hbar^2}$, $k_0 = \sqrt{2\mu_{cn}|\varepsilon_0|/\hbar^2}$, 
with $\mu_{cn}$ the core-neutron reduced mass. 
By defining $\hat l\equiv \sqrt{2l+1}$ for general angular quantum numbers, from the above  
we obtain
 {\small \begin{eqnarray}\label{electric}
 \frac{dB(E\lambda)}{d\varepsilon}&=&
\frac{\mu_{cn}}{\hbar^2 k}\sum_{\tilde{\j}}(2{\tilde{\j}}+1)|\mathcal{F}_{\lambda,{\tilde{\j}}}|^2,
\;\; {\rm with}\\
 \mathcal{F}_{\lambda,j}&\equiv& 
 \frac{1}{4\pi}Z_{\lambda}\hat\ell_0\hat\ell\hat\lambda^2\hat j_0\hat j
 (-1)^{\ell_0+\ell+s+j+j_0+I_c+{\tilde{\j}}}\nonumber\\
 &\times &
 \left(\begin{array}{ccccc}
    \ell & \lambda & \ell_0 \\
    0 & 0& 0 
 \end{array}
\right)
\left(\begin{array}{ccccc}
j & \lambda & j_0 \\
    0 & 0& 0 
 \end{array}
\right)
 \left \{\ \begin{array}{cccc}
 s &\ell_0 & j_0 \\
 \lambda & j & \ell
\end{array}
 \right\}\
 \nonumber\\
  &\times &
\left \{\ \begin{array}{cccc}
 I_c &j_0 & \tilde{\j}_0 \\
 \lambda & \tilde{\j} & j
\end{array}
 \right\}\
    \int_0^{\infty}dr\, u_{k_0\ell_0}^{\tilde{\j}_0}(r)r^{\lambda}u_{k\ell}^{\tilde{\j}}(r),\nonumber
\end{eqnarray}
}where  $u_{k_0\ell_0}^{\tilde{\j}_0}(r)$, and $u_{k\ell}^{\tilde{\j}}(r)$ are the ground-state and continuum radial 
wave functions.  The Eqs.~(\ref{CE})-(\ref{electric}) are indicating how the Coulomb breakup is being affected by 
the projectile structure.

\subsection{Computational details}
\begin{figure}[h]
\begin{center}
\hspace{-1mm}
\resizebox{75mm}{!}{\includegraphics{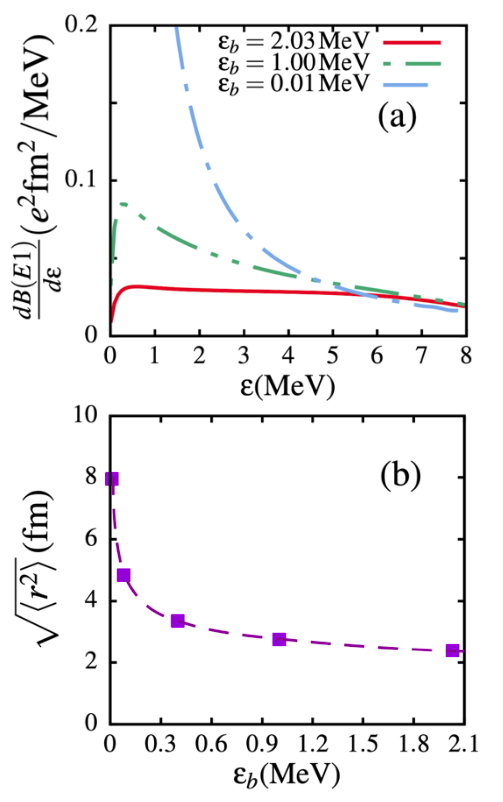}}
\end{center}
\caption{\label{f01} In panel (a), considering three different $^7$Li-n ground-state binding 
energies $\varepsilon_b$, it is shown how the derivative of the electric transition probability,
given by (\ref{electric}), 
varies with the excitation energy $\varepsilon$, for transitions from ground to continuum $s$- plus $d$-states. 
In panel (b),  the root-mean-square radii is shown as a function of the binding energy $\varepsilon_b$.}
\end{figure}

 The energies and corresponding wave functions which appear in the set of coupled differential equations 
 (\ref{coupled}), for the bound and continuum states of the $^7$Li+n system, are 
obtained by considering a two-body
Woods-Saxon potential as input, whose parameters are the same as in Ref.~\cite{Moro200}. 
The depth $V_0$ of the central part of the potential was adjusted to reproduce the ground and excited 
bound-state energies. 
These parameters are summarized in Table~\ref{table1}. 
\begin{table}[h]
\caption{\label{table1} Woods-Saxon potential parameters for the projectile (n$-^7$Li) ground and excited 
bound-state energies.}
\begin{tabular}{lllllllll}
\hline\hline $\tilde{\j}^{\pi}$
& \multicolumn{1}{c}{$V_0$} 
& \multicolumn{1}{c}{$r_0$} 
& \multicolumn{1}{c}{$a_0$} 
& \multicolumn{1}{c}{$V_{\rm SO}$} 
& \multicolumn{1}{c}{$r_{\rm SO}$} 
& \multicolumn{1}{c}{$a_{\rm SO}$} \\
& \multicolumn{1}{c}{(MeV)} 
&\multicolumn{1}{c}{(fm)}
&\multicolumn{1}{c}{(fm)} 
& \multicolumn{1}{c}{(MeV/fm$^{2}$)} 
& \multicolumn{1}{c}{(fm)} 
& \multicolumn{1}{c}{(fm)} \\
\hline
$2^+$
&\multicolumn{1}{c}{37.22} 
&\multicolumn{1}{c}{1.25} 
& \multicolumn{1}{c}{0.52} 
& \multicolumn{1}{c}{4.89}  
& \multicolumn{1}{c}{1.25} 
& \multicolumn{1}{c}{0.52} \\
$1^+$
&\multicolumn{1}{c}{46.65} 
&\multicolumn{1}{c}{ 1.25} 
& \multicolumn{1}{c}{0.52} 
& \multicolumn{1}{c}{4.89}  
& \multicolumn{1}{c}{1.25} 
& \multicolumn{1}{c}{0.52} \\
\hline\hline
\end{tabular}
\end{table}
Similarly, the other binding energies considered in this work are obtained by adjusting $V_0$.  The same 
ground-state potential 
parameters are adopted to calculate the corresponding continuum wave functions.
With these potential parameters, we first calculate the electric transition probability $B(E1)$ variation with the 
excitation energy $\varepsilon$, given by Eq.(\ref{electric}), corresponding to  the
transition from the ground-state to continuum $s$- plus $d$-states, 
for the binding energies $\varepsilon_b=0.01\,{\rm MeV}, 1.0\,{\rm MeV}$ and $2.03\,{\rm MeV}$. The results are 
shown in the upper panel of Fig.\ref{f01}.
One notices that $B(E1)$ varies substantially for $\varepsilon_b=0.01\,{\rm MeV}$ as compared with values 
obtained for larger $\varepsilon_b$.
These results highlight the strong dependence of  the Coulomb breakup on the
projectile internal structure, particularly in the asymptotic region.
 In this regard, it is also instructive to verify how the projectile root-mean-square 
radii $\sqrt{\langle r^2\rangle}$ vary with the  projectile 
ground-state binding energies. For that, we add the lower panel of Fig.~\ref{f01}, with 
the corresponding root-mean-square radii, obtained for the projectile ground-state $\Phi_{\beta_0}(\bm r)$.
As expected, the root-mean-square radii behavior is reflecting the large increasing of the wave function as the 
binding energy comes close to zero. Also, for $\varepsilon_b=2.033$\,MeV, we note that 
we obtain $\sqrt{\langle r^2\rangle}=2.39$\,fm, in very close agreement with the corresponding values reported 
in Refs.~\cite{2015Fan} and \cite{1988Tanihata} (respectively, $\sqrt{\langle r^2\rangle}=2.39\pm 0.05$\,fm and 
$\sqrt{\langle r^2\rangle}=2.37\pm 0.02$\,fm).

In order to evaluate the coupling matrix elements of Eq.~(\ref{coupled}), fragments-target optical potentials are needed.
The $^7$Li$+^{12}$C optical potential parameters were taken from Ref.\,\cite{Barioni100}, whereas 
the $^7$Li$+^{208}$Pb optical potential parameters were obtained from the $^7$Li global potential 
of Ref.~\cite{Cook300}, with the depth of the real part slightly modified to fit the elastic scattering experimental data.
For the $n-$target optical potentials, we adopted the global potential of Ref.~\cite{1969Green}.
The CDCC  limiting values of the model space parameters, used for the numerical solution of Eq.(\ref{coupled}),  
are listed in  Table~\ref{table2}, for the two targets we are considering,  $^{12}$C and $^{208}$Pb, where
$\ell_{\rm max}$ is the maximum angular momentum between $^7{\rm Li}$ and the neutron,  
$\lambda_{\rm max}$ is the maximum order of the potential multipole expansion, $\varepsilon_{\rm max}$ is the 
maximum bin energies, $r_{\rm max}$ is the maximum matching radius for bin potential integration, 
$L_{\rm max}$ is the maximum angular momentum of the relative c.m. motion, and $R_{\rm max}$ is the 
maximum matching radius of the integration for the coupled differential equations,  with $\Delta R$ the 
corresponding $R-$step size.
The reported main values are found to give enough converged results for $\varepsilon_b\ge $0.4 MeV. 
However, as we decrease the projectile binding energy, for
$\varepsilon_b\le 0.08$MeV, to guarantee enough good convergence and precision of the results we found
necessary to increase the maximum values for the projectile matching radius $r_{\rm max}$, for the
matching radius $R_{\rm max}$, and for the relative angular momentum of the c.m. motion,
$L_{\rm max}$, correspondingly to each of the target. These values for smaller $\varepsilon_b$ are shown
within parenthesis, below the respective values obtained for larger  $\varepsilon_b$.
The adopted bin widths were, $\Delta\varepsilon=0.5\,{\rm MeV}$, for $s$- and $p$-states, 
$\Delta\varepsilon=1.0\,{\rm MeV}$,  for $f$- and $d$-states and $\Delta\varepsilon=1.5\,{\rm MeV}$ for 
${\rm g}$-states.
{\small
\begin{table}[h]
\caption{\label{table2} 
Maximum model space parameters, for optimal numerical convergence of Eq.~(\ref{coupled}) 
for both $^{12}$C and $^{208}$Pb targets. The main reported values are for 
$\varepsilon_b\ge 0.4$MeV, with the corresponding ones within parenthesis for 
$\varepsilon_b\le 0.08$MeV.}
\begin{tabular}{cccccccc}
\hline\hline 
Target
&$\ell_{\rm max}$&$\lambda_{\rm max}$&$\varepsilon_{\rm max}$&$r_{\rm max}$&$L_{\rm max}$&
$R_{\rm max}$&$\Delta R$\\
&($\hbar$)           &{}                             &(MeV)                            &(fm)                &($\hbar$)         
&(fm)            &(fm)\\  
\hline
$^{12}{\rm C}$    &{3}                            &{3}                                & {6}                 & {80}                
& {300}        & {300} & {0.08} \\
                           &                                &                                    &                      & {(100)}            
                           & {(1000)}     & {(500)}& \\
$^{208}{\rm Pb}$ &{4} &{ 4} & {10} & {80}  & {1000} & {600} & {0.03}\\
                           & & & & {(100)}  & {(10000)} & {(1000)} & \\
\hline\hline
\end{tabular}
\end{table}}

\begin{figure}[t]
\begin{center}
\hspace{-0.8cm}
  \resizebox{75mm}{!}{\includegraphics{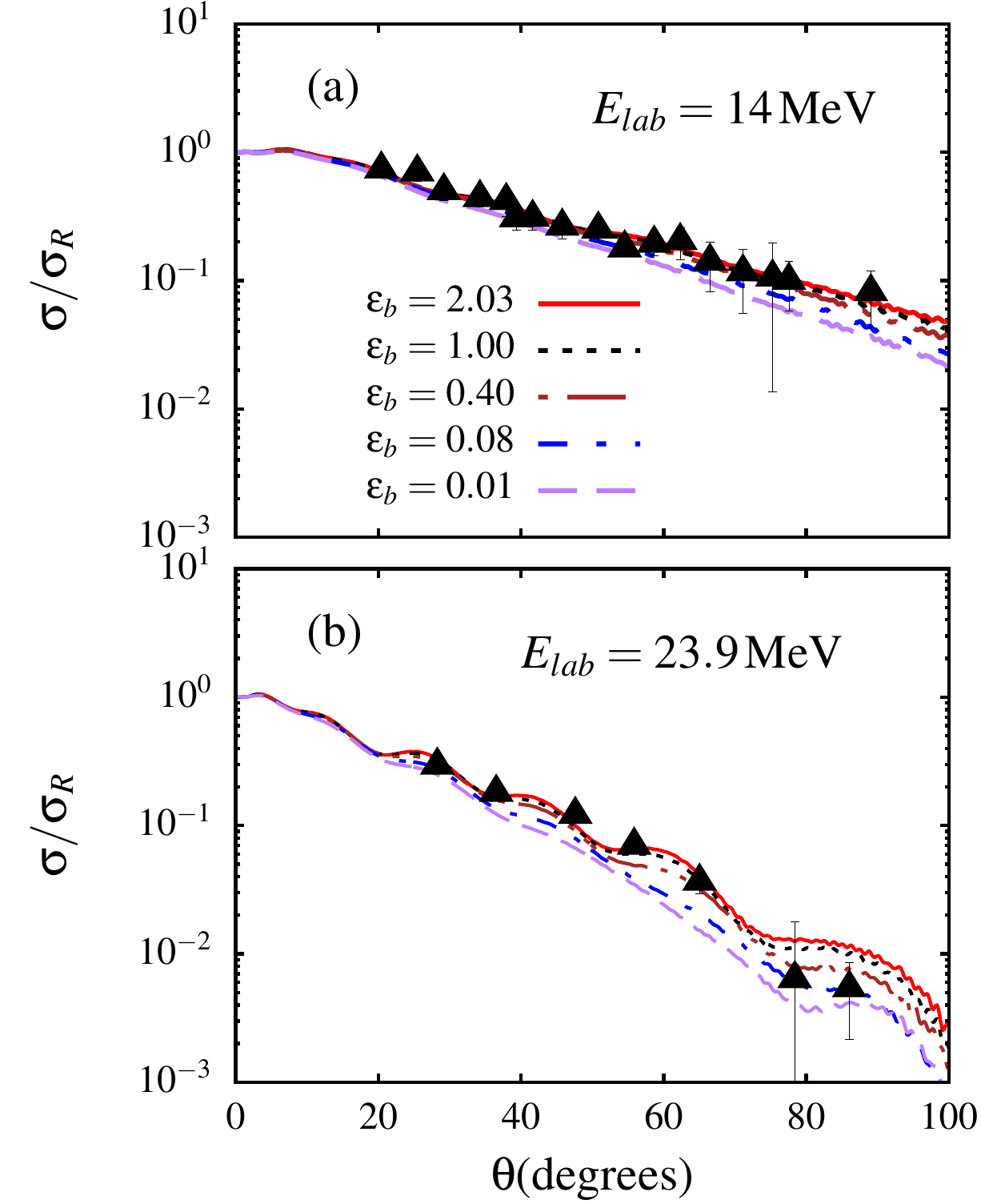}}
\end{center}
\caption{\label{f02}
$^8$Li+$^{12}$C elastic scattering cross sections for the incident energies, $E_{lab}=$14 MeV 
and $E_{lab}=$23.9 MeV. The model results are for different $^8$Li binding energies $\varepsilon_b$ 
(in MeV units), as indicated inside panel (a) for both panels. The available experimental data, converted
to Rutherford $\sigma_R$ units, are from Refs.~\cite{1993Becchetti} [panel (a)] and \cite{Barioni100} 
[panel (b)], as indicated in the database reported in Ref.~\cite{Jinr20}. 
}
\end{figure}
\begin{figure}[h]
\begin{center}
  \resizebox{75mm}{!}{\includegraphics{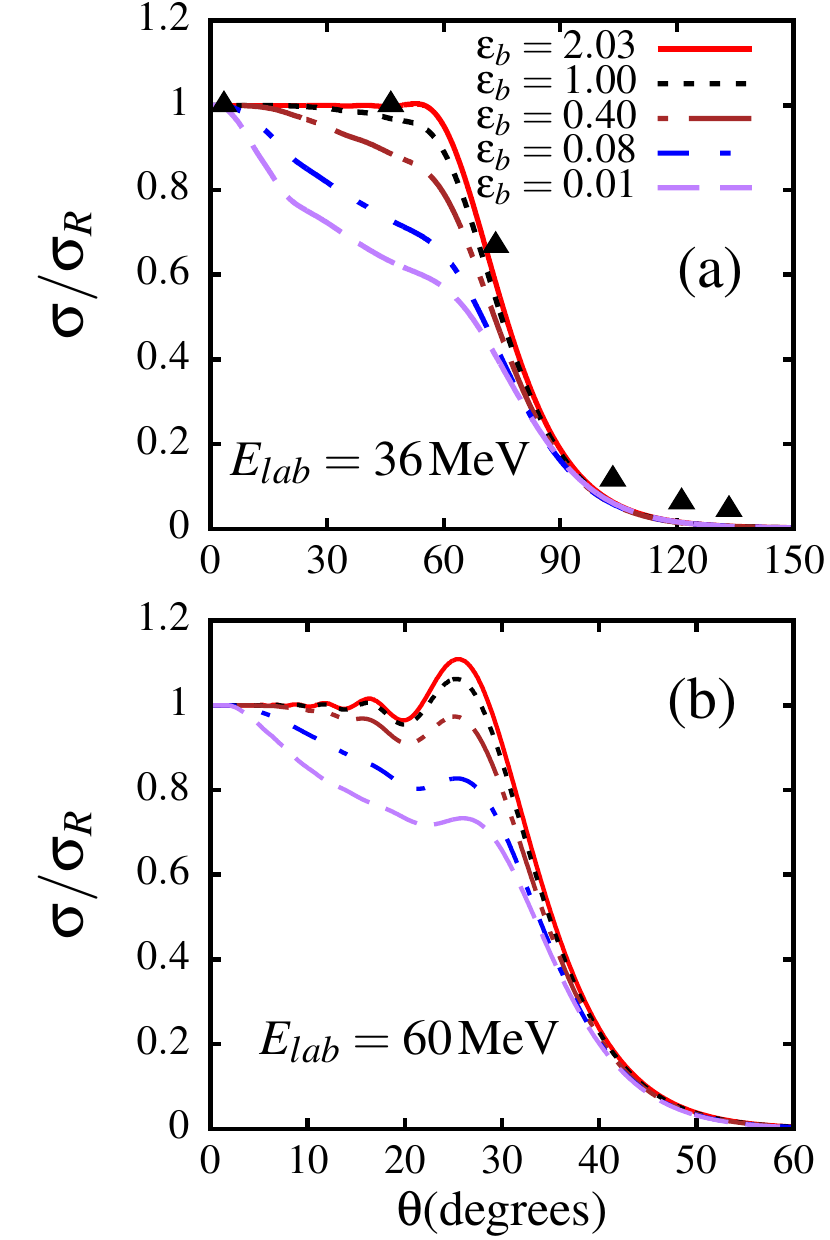}}
\end{center}
\caption{\label{f03}
$^8$Li+$^{208}$Pb elastic scattering cross sections (in units of the Rutherford $\sigma_R$), 
obtained for the incident energies $E_{lab}=$36 MeV [panel (a)] and $E_{lab}=$60 MeV [panel (b)].
As in Fig.~\ref{f02}, the results are for the same set of $\varepsilon_b$ (in MeV).
From Ref.~\cite{2002Kolata}, we included in (a) the closest available experimental data, which are for 
$E_{lab}=$30.6 MeV,  as indicated in the database reported in Ref.~\cite{Jinr20}.}
\end{figure}

\section{Results and Discussion}
\label{results}
\subsection{Elastic scattering cross sections}
\label{elastic}
We start the first part of this section by analyzing the dependence of the elastic scattering cross sections on the 
projectile ground-state binding energy.  These cross sections are displayed in Fig.~\ref{f02} for
 $^{12}{\rm C}$ target; and in Fig.~\ref{f03} for $^{208}{\rm Pb}$ target, considering two incident energies. In both
 the cases, we assume different values of $\varepsilon_b$,  from the experimental one down to 0.01 MeV.
In the case of $^{12}{\rm C}$ target, which is a nuclear-dominated reaction, from the results shown in Fig.~\ref{f02}
one can observe a weak dependence on $\varepsilon_b$ in the range $0.4\,{\rm MeV}\le\varepsilon_b\le 
2.03\,{\rm MeV}$,
for both incident energies, $E_{lab}=$14 MeV [panel (a)] and 24 MeV [panel (b)].
However,  it becomes relatively significant for $\varepsilon_b\le 0.08\,{\rm MeV}$ [see panel (a)]. Also shown in
Fig.~\ref{f02} is that the experimental data are well reproduced by the model for both incident energies.

For the Coulomb-dominated reaction with $^{208}{\rm Pb}$, the results given in  Fig.\,\ref{f03} for $E_{lab}=$36 
MeV (a) and 60 MeV (b)] are indicating strong dependence of the elastic scattering cross sections on all binding 
energies at forward angles (asymptotic region),  where the Coulomb breakup is particularly dominant. 
However, at backward angles (short distance), where the nuclear breakup is expected to provide meaningful effects,
the elastic cross sections become almost independent of the binding energy.

These results lead to a conclusion that, when the nuclear breakup is dominant or relatively significant, the effect of the 
binding energy on the elastic scattering cross section is rather small, whereas it is more pronounced when the 
Coulomb breakup is dominant.  Therefore, since a relatively significant effect for the $^8{\rm Li}+{}^{12}{\rm C}$ 
reaction is observed when $\varepsilon_b\le 0.08\,{\rm MeV}$,
it is possible that the $^8{\rm Li}+{}^{12}{\rm C}$ reaction is already dominated by 
the Coulomb breakup for $\varepsilon_b\le 0.08\,{\rm MeV}$.
As the binding energy decreases, the Coulomb breakup becomes dominant over its nuclear counterpart, as anticipated. 
It also follows that the probability of the projectile to fly on the outgoing trajectory unbroken decreases, diminishing the 
corresponding elastic scattering cross section. In the next section, we will look into this observation in more detail. 

\subsection{Breakup cross sections}\label{breakup}
\begin{figure}[h]\hspace{-5cm}
\begin{center}
\hspace{-5mm}
  \resizebox{85mm}{!}{
  \includegraphics{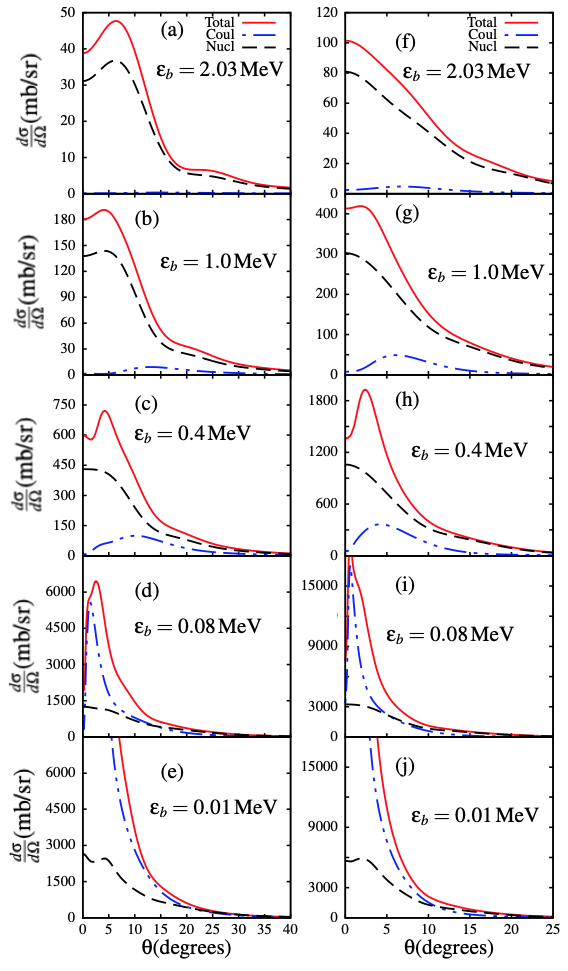}
  }
\end{center}
\caption{\label{f04}
For incident energies $E_{ lab}=14\,{\rm MeV}$ (left column) and $E_{ lab}=24\,{\rm MeV}$ (right column),
with fixed different $\varepsilon_b$ (shown inside the panels),
the $^8$Li$+^{12}$C angular distributions for the total, Coulomb and nuclear differential breakup cross sections 
$d\sigma/d\Omega$ (identified inside the upper panels) are shown as functions of the c.m. angle $\theta$.
}
\end{figure}
\begin{figure}[h]
\begin{center}
  \resizebox{85mm}{!}{
  \includegraphics{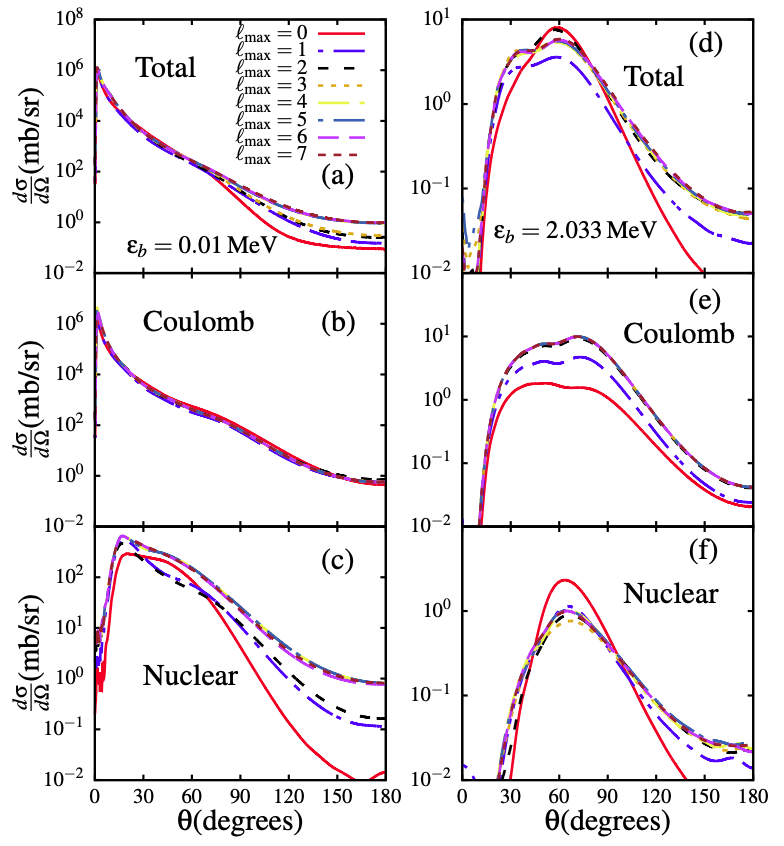}}
\end{center}
\caption{\label{f05}
Convergence sample results for the $^8$Li$+^{208}$Pb, 
total (upper frames), Coulomb (middle frames) and nuclear (bottom frames)
 breakup angular distributions, $d\sigma/d\Omega$, at $E_{ lab}=36\,{\rm MeV}$, considering 
 different maximum projectile internal angular momenta $\ell_{\rm max}$ (indicated in the upper-left frame).
The left set [(a)-(c)] is for $\varepsilon_b=0.01$MeV, with the right set [(d)-(f)] for $\varepsilon_b=2.03$MeV.
}
\end{figure}
\begin{figure}[h]
\begin{center}
  \resizebox{80mm}{!}{
  \includegraphics{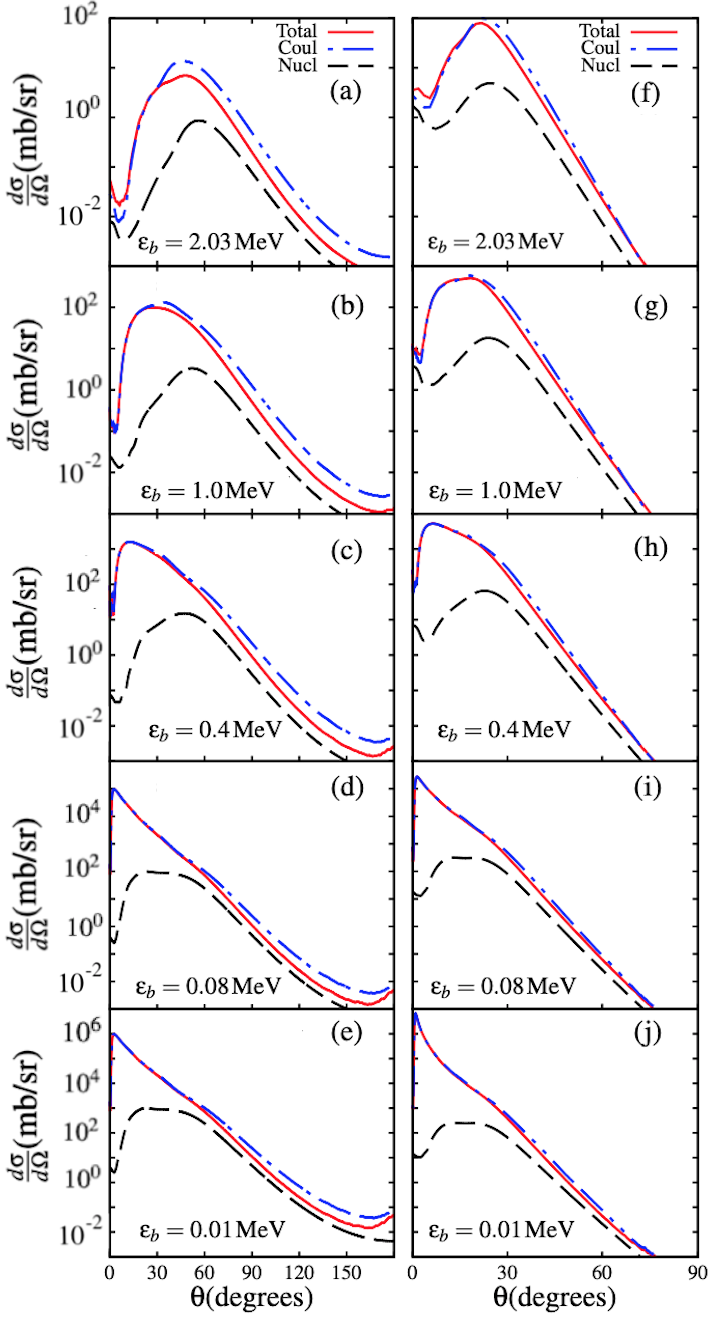}
  }
\end{center}
\caption{\label{f06}
For incident energies $E_{ lab}=36\,{\rm MeV}$ (left column) and $E_{ lab}=60\,{\rm MeV}$ 
(right column), with fixed different $\varepsilon_b$ (shown inside the panels),
the $^8{\rm Li}+{}^{208}{\rm Pb}$ angular distributions for the total, Coulomb and nuclear 
$d\sigma/d\Omega$ (identified in the upper panels) are shown as functions of the c.m. angle $\theta$.
}
\end{figure}

The differential total, Coulomb and nuclear breakup cross sections, for the $^{12}{\rm C}$ target, 
are depicted in Fig.~\ref{f04},  for $E_{lab}=$14 MeV [(a)-(e) panels] and $E_{lab}=$24 MeV [(f)-(j) panels].
As anticipated, in the case of nuclear-dominated reactions, for both incident energies, 
$d\sigma_{nucl}/d\Omega$ $\simeq d\sigma_{tot}/d\Omega $ $\gg d\sigma_{Coul}/d\Omega$ 
as $\varepsilon_b\to 2.03\,{\rm MeV}$, with $d\sigma_{Coul}/d\Omega\to 0$.
However, it is interesting to notice that as $\varepsilon_b$ decreases, the Coulomb breakup increases rapidly, 
such that for $\varepsilon_b\to 0.01\,{\rm MeV}$,
$d\sigma_{nucl}/d\Omega$
$\ll d\sigma_{Coul}/d\Omega$ $\simeq d\sigma_{tot}/d\Omega$ 
at forward angles, for both incident energies. 
On the light of these results it follows that as the binding energy further decreases, the Coulomb breakup becomes 
more relevant, and comparable with the total breakup even in such a naturally nuclear-dominated reaction.
This can be attributed to the fact that the breakup becomes more peripheral as $\varepsilon_b$ decreases, where 
only Coulomb forces are available.
Hence, the importance of the Coulomb breakup in this case relies mainly on the long-range behavior of the Coulomb forces,
and on its direct dependence on the electromagnetic transition matrix elements,
in agreement with our assessment in Sect.~\ref{elastic}. 
Furthermore, these results show that the ``nuclear-dominated reaction" concept may 
be relative to the projectile binding energy.

As the projectile binding energy varies from 2.03 MeV down to 0.01 MeV, one may wonder how relevant 
higher-order partial-waves ($\ell$)
are in the breakup process for such very low binding energy, particularly for heavy targets.
In order to verify the importance of higher-order partial-waves in this case, we performed a convergence test of the total,
Coulomb and nuclear
differential breakup cross sections for $^{208}$Pb target at $E_{lab}=36$ MeV. The different breakup cross sections
are shown in  Fig.~\ref{f05},
as functions of the c.m. angle $\theta$, for different maximum projectile internal angular momenta
  $\ell_{\rm max}$, and only for $\varepsilon_b=0.01$\,MeV and 2.03\,MeV binding energies.
  As evidenced by the results in this figure, there is no meaningful difference between $\ell_{\rm max}=4$ and 
  $\ell_{\rm max}=7$, regardless the binding energy. 
  This implies that, by reducing the ground-state binding energy,
  the convergence of the breakup cross sections is not affected, in respect to 
  the maximum core-neutron orbital angular momentum $\ell_{\rm max}$.
  
 In Fig.\ref{f06}, displays, the total, Coulomb and 
 nuclear breakup angular cross-section distributions as functions of the c.m. angle $\theta$, for the different binding 
 energies $\varepsilon_b$, for $^8$Li$+^{208}$Pb reaction. We first observe that as $\varepsilon_b$ 
 decreases, the peaks of $d\sigma_{tot}/d\Omega$ and  $ d\sigma_{Coul}/d\Omega$ are
 shifted to forward angles. In fact, for $\varepsilon_b\le$ 0.08 MeV, the peaks are located close to zero degree. 
 This is a clear display of the peripheral nature of the breakup process as $\varepsilon_b$ decreases.  A careful look 
 at this figure also indicates that as $\varepsilon_b$ decreases, even the peak of  $d\sigma_{nucl}/d\Omega$ is shifted
 to forward angles, which may suggest that even the nuclear breakup process becomes peripheral as 
 $\varepsilon_b\to$ 0.01 MeV.  The peripherality of the nuclear breakup in this case, can be understood by
considering the function $\delta_R(\varepsilon_b)$, which appears in Eq.(\ref{potnucl}).  The nuclear breakup 
dynamics require  that $\delta_R(\varepsilon_b)\to 0$, as $\varepsilon_b$ increases, implying that $R_n\to R_0$, 
due to the short-range nature of nuclear forces.  However, as $\varepsilon_b\to 0$, $\delta_R(\varepsilon_b)$ 
increases and so does $R_n$, leading to a significant nuclear effect in the peripheral region. 
Therefore, the function $\delta_R(\varepsilon_b)$ is introduced to take into account the 
well-known effect which occurs in weakly-bound systems, as in halo nuclei, in which the nuclear forces are
stretched beyond the usual range.

Quantitatively, since this $^8$Li$+^{208}$Pb reaction is Coulomb-dominated, we observe that at forward angles 
 both $d\sigma_{tot}/d\Omega$ and $d\sigma_{Coul}/d\Omega$ are substantially larger than 
 $d\sigma_{nucl}/d\Omega$ (about three orders of magnitude as $\varepsilon_b$ decreases). A further inspection 
 of this figure shows that for $E_{lab}=60$ MeV, we notice that the total and Coulomb breakup
 cross sections are more similar compared to $E_{lab}=36$\,MeV,
 with the difference coming from the competition between the nuclear and Coulomb interactions
 above the barrier
 (for a discussion on the role of the diagonal Coulomb interaction, see also Ref.~\cite{MukeruPRC2020}). 
\begin{figure}[!t]
\begin{center}
  \resizebox{70mm}{!}{\includegraphics{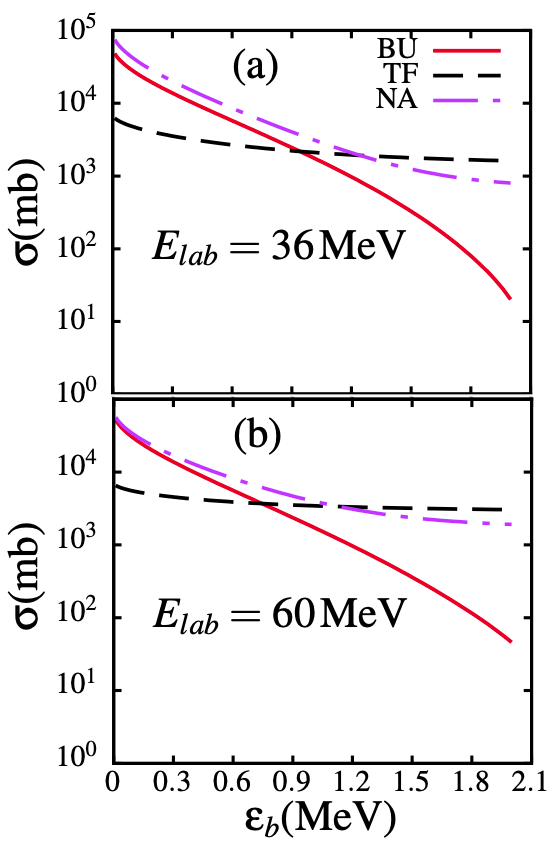}}
\end{center}
\caption{\label{f07} For the $^8$Li$+^{208}$Pb reaction, by considering 
$E_{lab}=$36 MeV (a) and 60 MeV (b), it is shown the integrated breakup cross sections 
(BU) (when both $W_{ct}^{nucl}$ and $W_{nt}^{nucl}$ are contributing), 
the breakup cross section without absorption (NA) (when $W_{ct}^{nucl}=W_{nt}^{nucl}= 0$),
and the total fusion cross section (TF), as functions of the projectile binding energy $\varepsilon_b$.}
\end{figure}

In order to better elucidate the importance of the nuclear absorption in the breakup process,
we present in Fig.~\ref{f07}, for the $^8$Li+$^{208}$Pb reaction, the integrated total breakup cross section  
as well as the  total fusion cross sections as functions of $\varepsilon_b$.
In this regard, we are extending a previous analysis done for this reaction in Ref.~\cite{2020Mukeru}, in which 
the total fusion cross sections are shown as functions of the incident energy for different projectile binding energies. 
The breakup cross section obtained in the presence of nuclear absorption  (i.e., $W_{ct}^{nucl}\ne 0$, 
$W_{nt}^{nucl}\ne 0$),  are indicated by the label ``BU". The breakup cross section obtained in the
 absence of nuclear absorption (i.e, $W_{ct}=W_{nt}=0$), 
are indicated by ``NA".  The total fusion cross section is  labeled  as ``TF''.  By observing this figure, it follows that,
 as $\varepsilon_b\to 2.03$ MeV, the nuclear absorption contributes to largely reduce the breakup cross section 
 about one order magnitude in the log-scale. However, we observe that the nuclear absorption plays a minor role 
 on the breakup cross section for smaller binding energies, being  negligible  for $\varepsilon_b\to 0.01$ MeV, 
 in particular at $E_{lab}=60$\,MeV.  In this case, $\sigma_{\rm NA}\simeq\sigma_{\rm BU}\gg\sigma_{\rm TF}$, 
 ({ where $\sigma_{\rm BU}$ is the breakup cross section
 followed by fragments absorption, and $\sigma_{\rm NA}$ is the breakup cross section without fragments absorption 
 after breakup}). 
A larger breakup cross section over the total fusion cross section 
 can be understood as due to the fact that, when the breakup occurs 
  where classically the trajectory is far away from the target, 
  the projectile fragments have no easy access to the absorption region, thus significantly reducing the flux that 
  contributes to the fusion cross section. However, as expected, as $\varepsilon_b\to 2.03$ MeV, where the breakup process 
 occurs closer to the target, where the probability for the projectile fragments to survive absorption 
 is significantly reduced, we observe that $\sigma_{\rm BU}\ll \sigma_{\rm NA}<\sigma_{\rm TF}$. 
A weak dependence of the total fusion cross section on the binding energy compared to the breakup cross 
section is also observed.  The energy region well above the Coulomb barrier is particularly
dominated by the complete fusion process. As shown in Ref.~\cite{Lei100}, the complete fusion 
cross section is insignificantly dependent on the projectile $\varepsilon_b$ for $^7$Li$+^{209}$Bi reaction.
We believe that these observations would be valid for any loosely bound projectile, and hence there
is nothing unusual in the breakup of the $^8{\rm Li}$ nucleus. 

Concerning our approach to total fusion (TF) and absorption, let us clarify that:
In the standard CDCC method, the optical potentials are chosen to describe the elastic scattering of the 
fragments by the target. So, their imaginary parts account for the absorption to fusion and other direct 
channels (surface reactions). Nevertheless, as direct reaction cross sections are expected to be small 
for the interactions between fragments and targets selected in this work,
the TF cross section provides the major contribution to this absorption.

\begin{figure}[h]
\begin{center}
  \resizebox{70mm}{!}{\includegraphics{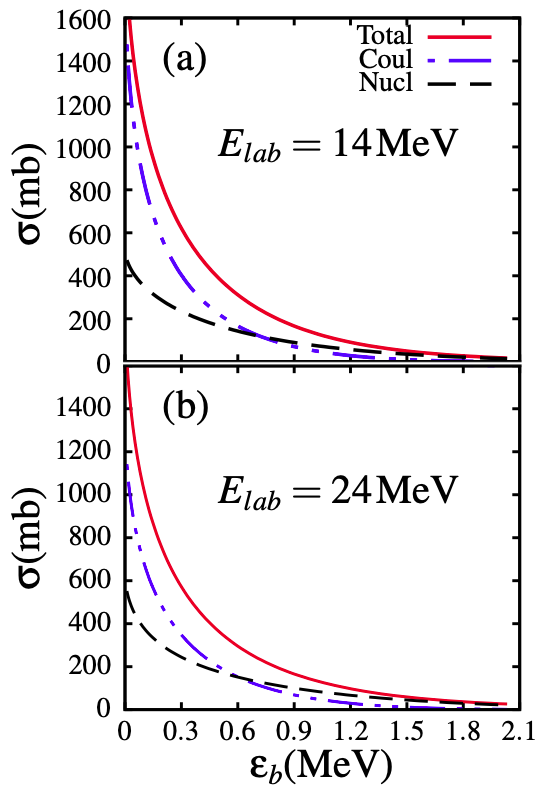}}
\end{center}
\caption{\label{f08}
For the $^8$Li+$^{12}$C breakup reaction, the angular-integrated total, Coulomb and 
nuclear breakup cross sections are given as functions of the projectile binding energy $\varepsilon_b$,
for the incident energies $E_{lab}=$14 MeV (a) and  24 MeV (b).}
\end{figure}
\begin{figure}[h]
\begin{center}\hspace{-.5cm}
\resizebox{80mm}{!}{
  \includegraphics{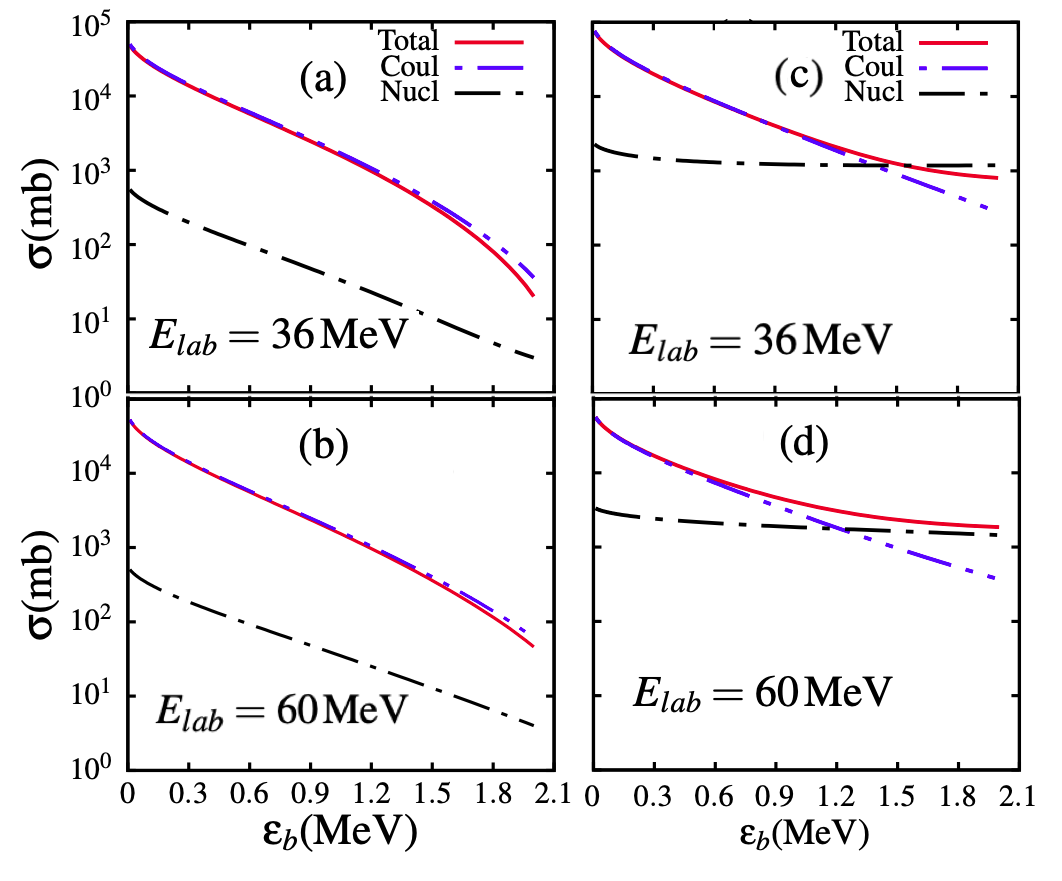}}
\end{center}
\caption{\label{f09} 
The angular-integrated total, Coulomb and nuclear breakup cross sections are given for the
$^8$Li+$^{208}$Pb breakup reaction as functions of  $\varepsilon_b$, with nuclear absorption
in the panels (a) and (b); and without absorption in the panels (c) and (d).  As indicated, the incident 
energies are $E_{lab}=$36 MeV [panels (a) and (c)] and 60 MeV [panels (b) and (d)]. 
}\end{figure}

For a better quantitative assessment of these results, we consider the integrated total 
($\sigma_{tot}$), Coulomb ($\sigma_{ Coul}$), and nuclear ($\sigma_{nucl}$)
breakup cross sections, which are 
displayed as functions of $\varepsilon_b$ in Fig.~\ref{f08} (for the $^{12}{\rm C}$ target), and
in Fig.~\ref{f09} (for the $^{208}{\rm Pb}$ target).
The results in both figures confirm the conclusions already drawn from Figs.\ref{f04} and \ref{f06}. For example, both
panels of Fig.\ref{f08}, show that as $\varepsilon_b\to$ 0.01 MeV, $\sigma_{Coul}> \sigma_{nucl}$ 
($\sigma_{Coul} \simeq\sigma_{tot}$), whereas
$\sigma_{Coul} < \sigma_{nucl}\simeq\sigma_{tot}$ ($\sigma_{nucl}\simeq\sigma_{tot}$) as $\varepsilon_b\to 2.03$ MeV. 
For $^{208}{\rm Pb}$ target, the results are shown 
in the presence of nuclear absorption. When the nuclear absorption is taken into account [panels (a) and (b)], we notice that
$\sigma_{coul}\simeq\sigma_{tot}\gg \sigma_{nucl}$ and this is independent of $\varepsilon_b$.
In the absence of the nuclear absorption [panels (c) and (d)], while $\sigma_{coul}\simeq\sigma_{tot}\gg \sigma_{nucl}$ 
remains valid for $\varepsilon_b\to 0.01$ MeV, it is noticed that $\sigma_{tot}\simeq \sigma_{nucl}>\sigma_{Coul}$, 
for $\varepsilon_b\to 2.03$ MeV, which further highlights the importance of the nuclear absorption for large binding 
energies.  The results in this figure further support the fact that strong nuclear absorption in the inner region is the 
main  factor that dictates the importance of the Coulomb breakup cross section over its nuclear counterparts.   
\begin{table*}[t] 
\caption{\label{table3} Coulomb, nuclear and interference cross-sections for the $^8{\rm Li}+{}^{12}{\rm C}$ 
and $^8$Li$+^{208}$Pb,  considering n$-^7$Li binding energies $\varepsilon_b=$0.01 MeV
and 2.03 MeV. For each target, we present our results, in terms of ratios, for two colliding energies.
For $^{208}$Pb target, with no-nuclear absorption (NA) the results are shown within parenthesis
below the ones with absorption. }
\begin{center}
\begin{tabular}{cc|ccccc|ccccc}
\hline\hline
Target  &${E_{ lab}}$ &&$\varepsilon_b=$&2.03&MeV&&&$\varepsilon_b=$&0.01&MeV&\\
&(MeV)& 
$\frac{\sigma_{Coul}}{\sigma_{tot}}$   & 
$\frac{\sigma_{nucl}}{\sigma_{tot}}$    &
$\frac{\sigma_{Coul}}{\sigma_{nucl}}$ &
$\frac{\sigma_{int}}{\sigma_{nucl}}$    &
$\frac{\sigma_{int}}{\sigma_{tot}}$    &
$\frac{\sigma_{Coul}}{\sigma_{tot}}$  & 
$\frac{\sigma_{nucl}}{\sigma_{tot}}$   &
$\frac{\sigma_{Coul}}{\sigma_{nucl}}$ &
$\frac{\sigma_{int}}{\sigma_{nucl}}$    &
$\frac{\sigma_{int}}{\sigma_{tot}}$ 
               \\ \hline\hline
$^{12}{\rm C}$     & 14 &0.024&0.824&0.029&0.186&0.153  &0.836&0.268&3.123&-0.387& -0.104\\
                            & 24 &0.048&0.808&0.059&0.190&0.154  &0.701&0.339&2.069&-0.118& -0.040\\ 
                            \hline                           
$^{208}{\rm Pb}$ &36&1.800&0.150&12.00&-6.333&-0.950 &1.033&0.012&90.16&-3.850& -0.044\\
                      &&(0.344) &(1.481)&(0.232)&(-0.557)&(-0.825)&(1.032)&(0.029)&(33.97)&(-2.061)&(-0.063)\\
                          &60&1.326&0.087&15.25&-4.750&-0.413&1.015&0.010&104.6&-2.540&-0.025\\
                    &&(0.198)&(0.783)&(0.253)&(0.025)&(0.019) & (1.000)&(0.059)&(17.03)&(-0.991)&(-0.058) \\
               \hline\hline
\end{tabular}
\end{center}
    \end{table*}
In Table~\ref{table3}, we provide more quantitative results, given as fractions from $\sigma_{tot}$ and 
$\sigma_{nucl}$,  reflecting the competition between the different cross sections, by selecting the two 
limiting binding energies we are studying, i.e., $\varepsilon_b=$0.01 MeV and $\varepsilon_b=$2.03 MeV. 
We are also including $\sigma_{int}$, as defined by 
\begin{eqnarray}\label{sigint}
  \sigma_{int}=\sigma_{tot}-(\sigma_{Coul}+\sigma_{nucl}),
\end{eqnarray}
which we naively regard as the Coulomb-nuclear interference and that will be discussed in the next subsection.
From this table,  it becomes evident that, when $\varepsilon_b$ decreases, $\sigma_{Coul}$ (approaching to
$\sigma_{tot}$) becomes substantially larger than $\sigma_{nucl}$.  
Also, for the light $^{12}$C target, 
at $E_{lab}=14$\,MeV and 24\,MeV,  we note that  $\sigma_{Coul}/\sigma_{nucl}$ rapidly grows, when varying
$\varepsilon_b$ from 2.03 MeV down to 0.01 MeV. As shown, in this energy interval, 
$\sigma_{Coul}/\sigma_{nucl}$ increases from 0.03 to 3.12 for 14 MeV, and from 0.06 to 2.10 for 24 MeV. 
This indicates that, as the binding energy decreases, the $^8$Li$+^{12}$C reaction 
becomes like a ``Coulomb-dominated reaction",  with the emergence of a long-range behavior.
Moreover, with the heavy target at $E_{lab}=$36 MeV, in the presence of nuclear absorption, for 
$\varepsilon_b=$2.03 MeV,  $\sigma_{coul}/\sigma_{nucl}=12$, whereas
$\sigma_{coul}/\sigma_{nucl}\simeq 90$ for $\varepsilon_b=$ 0.01 MeV. 
It is noticed in this case that this ratio is substantially affected in the absence of nuclear absorption 
(NA),  becoming $\sigma_{coul}/\sigma_{nucl}\simeq 0.06$ ($\varepsilon_b=2.03$ MeV), and 
$\sigma_{coul}/\sigma_{nucl}\simeq 34$ ($\varepsilon_b=0.01$ MeV).
 \begin{figure}[!h]
\begin{center}
  \resizebox{70mm}{!}{\includegraphics{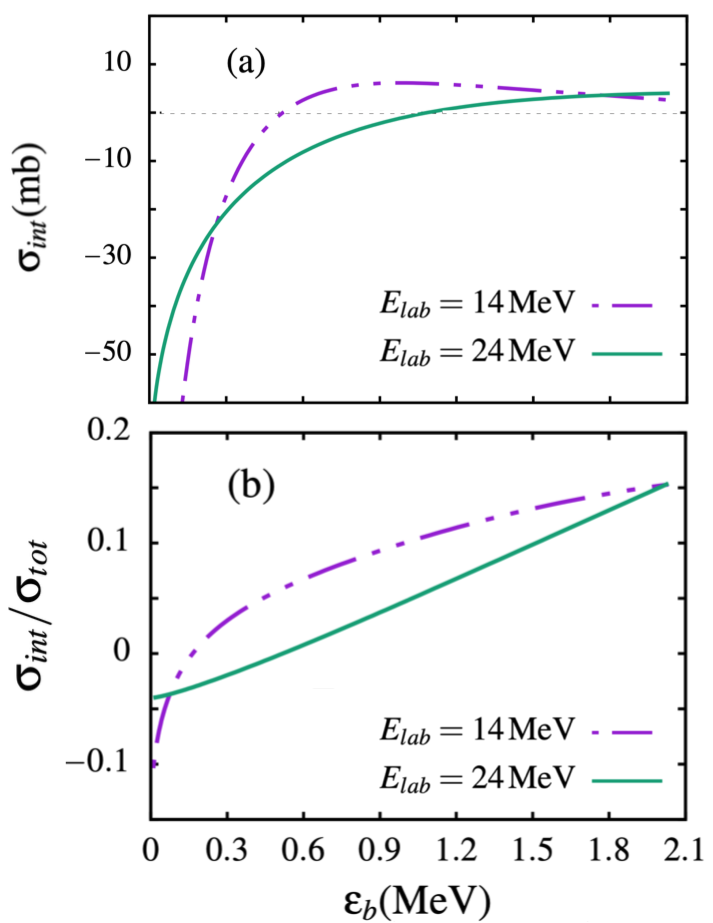}}
\end{center}
\caption{\label{f10} The $^8$Li+$^{12}$C integrated Coulomb-nuclear interference 
$\sigma_{int}$ [panel (a)], given by (\ref{sigint}), with the respective ratio $\sigma_{int}/\sigma_{tot}$ [panel (b)], 
are shown as functions of $\varepsilon_b$, for the colliding energies $E_{\rm lab}=$14 and 24 MeV.
}
 \end{figure}
\begin{figure*}[t]
\begin{center} 
  \resizebox{140mm}{!}
  {\includegraphics{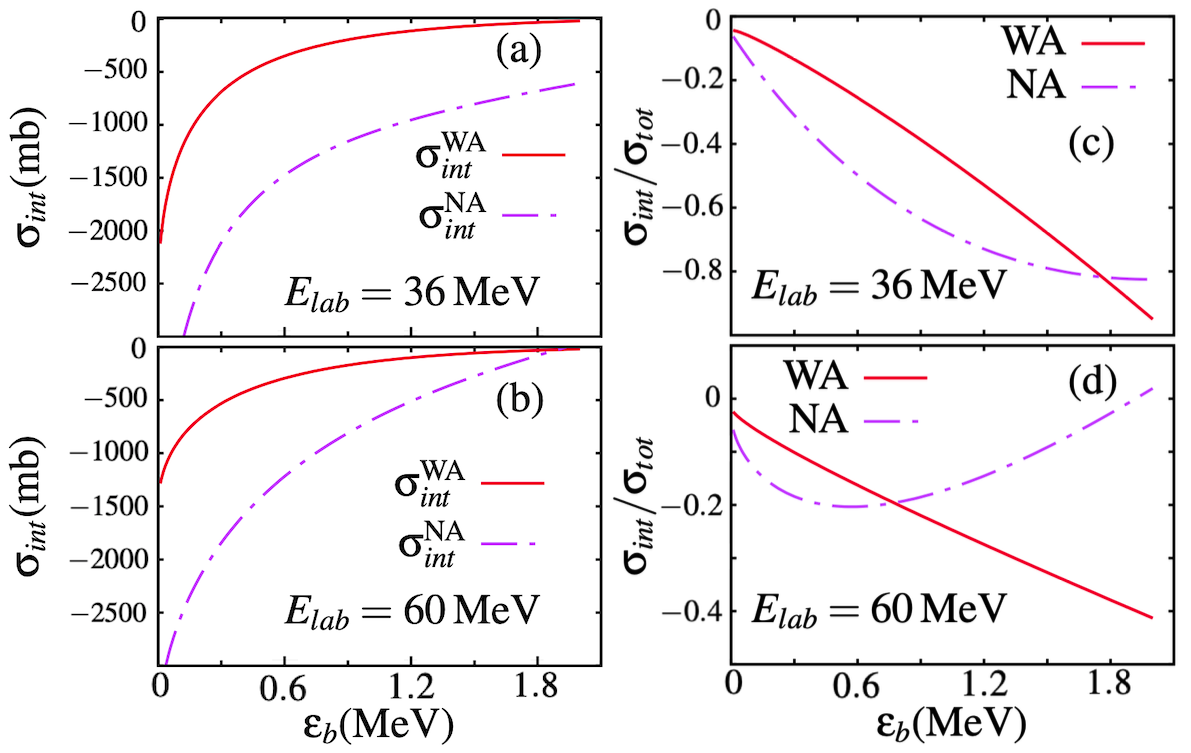}}
\end{center}
\caption{\label{f11} The $^8{\rm Li}+{}^{208}{\rm Pb}$ integrated Coulomb-nuclear interference $\sigma_{int}$
[panels (a) and (b)], given by (\ref{sigint}), with their ratios $\sigma_{int}/\sigma_{tot}$ [panels (c) and (d)],
are shown as functions of $\varepsilon_b$,  for $E_{\rm lab}=$36 and 60 MeV (upper and lower frames, respectively).
 $\sigma_{int}^{\rm WA}$ (solid lines) denotes the interference when the breakup is followed by nuclear absorption,
 with $\sigma_{int}^{\rm NA}$ (dot-dashed lines) denoting interference with no nuclear absorption.}
\end{figure*}

\subsection{Coulomb-nuclear interference}
\label{interf}
It is well-known that the incoherent sum of the Coulomb and nuclear breakup cross section 
$(\sigma_{Coul}+\sigma_{nucl})$ is always different from their coherent sum, $\sigma_{tot}$, due to the 
Coulomb-nuclear interference effect. To assess  this effect in the context of very weak ground-state 
binding energy,  we consider $\sigma_{int}$ as defined in Eq.~(\ref{sigint}) to estimate the 
Coulomb-nuclear interference.  Given that, for the two limiting binding energies, the quantitative results for
$\sigma_{int}$ are already furnished in Table~\ref{table3} as ratios with respect to $\sigma_{tot}$ and 
$\sigma_{nucl}$. In Figs.~\ref{f10} and \ref{f11} (respectively, for $^{12}$C and $^{208}$Pb targets),
we provide the exact $\sigma_{int}$ behaviors, together with their respective ratios
$\sigma_{int}/\sigma_{tot}$, as functions of $\varepsilon_b$, in a way to clarify that
the differences between $\sigma_{tot}$ and $(\sigma_{Coul}+\sigma_{nucl})$ are quite large in both
the cases, with the amount varying with $E_{lab}$ ($|\sigma_{int}|$ decreasing with increasing $E_{lab}$).
 The Coulomb-nuclear interference is strongly dependent on $\varepsilon_b$.  As one can notice,
 it appears to increase as $\varepsilon_b$ decreases, and becomes quite small as $\varepsilon_b\to 2.03$\,MeV.
 
 For the $^8$Li+$^{208}$Pb reaction, nuclear absorption which is already shown to reduce the breakup cross section
   (Fig.\ref{f07}), is expected to be more relevant on the Coulomb-nuclear interference.  The Coulomb-nuclear
   interference obtained when the breakup is followed by nuclear absorption (i.e., $W_{ct}^{nucl}\ne 0, W_{nt}^{nucl}\ne 0$),
   is denoted by $\sigma_{int}^{\rm WA}$ (WA standing for ``with absorption''),  and by $\sigma_{int}^{\rm NA}$
   the Coulomb-nuclear interference obtained when $W_{ct}^{nucl}=W_{nt}^{nucl}=0$. Therefore, in order to assess the relevance of the nuclear
   absorption on this interference, we compare $\sigma_{int}^{\rm WA}$ with $\sigma_{int}^{\rm NA}$. The results are presented in Fig.\ref{f11}.
 In this figure, the panels (a) and (b) are for the exact $\sigma_{int}$ results, whereas in
panels (c) and (d) we have the respective ratios $\sigma_{int}/\sigma_{tot}$. The upper panels are 
for $E_{lab}=36$ MeV, and the lower panels for $E_{lab}=60$ MeV.
The absorption contribution to $\sigma_{int}$ is verified by the observed 
difference $|\sigma_{int}^{\rm NA}-\sigma_{int}^{\rm WA}|$, which are clearly shown for both
$E_{lab}$ energies, as $\varepsilon_b$ varies.

Besides the fact that the Coulomb-nuclear interference is shown to be larger in the very 
small binding energy limits,  such larger values 
may also be influenced by the large magnitudes of the total and Coulomb breakup cross sections,
which are shown in Figs.~\ref{f09} and \ref{f10}. However, as verified from 
Table~\ref{table3}, the ratios $\sigma_{Coul}/\sigma_{tot}$ for the smaller binding 
are enough deviating from one (when full 
absorption is considered, in the $^{208}$Pb case). 
Consistently, we also noticed from the results given in Table~\ref{table3}, 
that $\sigma_{int}/\sigma_{tot}$ is larger for $\varepsilon_b=2.03$\,MeV when
we have the usual cross section values with absorption.
Further investigation may be required to clarify the $\varepsilon_b$ dependence of 
Coulomb-nuclear interference, in support to the actual results that are shown an overall significant effect
of nuclear absorption.

 In the case of such weakly-bound projectiles, a better understanding of the function $\delta_R(\varepsilon_b)$, 
 which appears in Eq.(\ref{potnucl}), could shed more light on the complexity of the Coulomb-nuclear interference. 
 In such cases, $R_n$ can significantly deviate from $R_0$,
 since the nuclear breakup dynamics requires  that $\delta_R(\varepsilon_b)\to 0$ for
 larger values of $\varepsilon_b$. 
Particularly, the main characteristics of this function could show up in a study with charged projectiles, considering 
that strong Coulomb/nuclear interference has been observed for the reaction of proton halo $^8$B with
$^{58}$Ni target~\cite{2001Tostevin,2002Margueron,Tarutina10,2009Lubian}, in which
we have a very weakly-bound projectile with breakup threshold of 0.137 MeV.

\section{Conclusion}
\label{conclusion}
We have presented a study on the breakup of the weakly-bound  $^8$Li  (n$-^7$Li) projectile on light
and heavy target masses, namely, $^{12}$C and $^{208}$Pb.
Our main objective was to investigate the dependence of the total, Coulomb and nuclear breakup cross sections,
on the $^8$Li ground-state binding energy $\varepsilon_b$, in order to study the peripherality of the total, Coulomb
  and nuclear breakup processes, which are associated to the weaker binding energy of the projectile.
To this end, apart from the experimen\-tally-known ground-state binding energy of the n$-^7$Li system,
we artificially considered four other binding energies, below the experimental value, down to
  $\varepsilon_b=0.01$\,MeV, which is much smaller than the experimental value, $\varepsilon_b=2.03$\,MeV.
 From our analysis it is shown that the total, Coulomb and nuclear breakup processes become peripheral as
$\varepsilon_b\to 0.01$\,MeV, regardless the target mass. 
We argue that the peripherality of the nuclear breakup in this case 
is primarily related to the spacial extension of the corresponding ground-state wave function, which is related to  
weaker binding energy. 
The peripheral region is determined by the range of the nuclear forces $R_0$, and the corresponding 
extension of the ground-state wave function, which is associated to a function $\delta_R(\varepsilon_b)$, 
expressed by $R_n$ defined in  Eq.~(\ref{potnucl}). By taking into account the fact that 
close to the $n-$core $\varepsilon_b\to 0$ limit, a long-range interaction is expected  
to emerge between projectile and target (similar as for three-body halo-nuclei systems~\cite{2012Frederico}),
the size of the associated wave function will increase significantly in this limit.
So, a detailed investigation of this function $\delta_R(\varepsilon_b)$ (which should go to zero by increasing 
$\varepsilon_b$) can shed more light into the dynamics of nuclear breakups induced by 
loosely bound projectiles.  
It is also noticed that the variation of $\varepsilon_b$ strongly affects the Coulomb breakup, as 
compared to the nuclear breakup, such that as $\varepsilon_b\to$ 0.01 MeV, the Coulomb breakup 
becomes dominant even for the $^{12}{\rm C}$ target, which is known to be naturally dominated by 
nuclear breakup.  Therefore, in view of this binding-energy dependence, 
one may infer that the expression ``naturally-dominated by nuclear breakup'' may be relative to the projectile 
binding energy.  It is also verified that the nuclear absorption has an insignificant effect on the total and nuclear 
breakup cross sections when decreasing the binding energy to small binding such as $\varepsilon_b\to$ 0.01 MeV. 
In this small binding energy region, we found that the total breakup cross section is larger than the calculated 
total fusion cross section, while as expected, the opposite is observed as $\varepsilon_b\to$ 2.03\,MeV.
\section*{Acknowledgements}
We thank T. Frederico, B.V. Carlson and L. F. Canto for useful discussions.
B.M. is also grateful to the South American Institute of Fundamental Research 
(ICTP-SAIFR) for local facilities. 
For partial support, we also thank 
Conselho Nacional de Desenvolvimento Cient\'\i fico e Tecnol\'ogico [INCT-FNA Proc.464898/2014-5 
(LT and JL), Proc. 304469/2019-0(LT) and Proc. 306652/2017-0(JL)], and Funda\c c\~ao de Amparo \`a Pesquisa do Estado de S\~ao Paulo 
[Projs. 2017/05660-0(LT)].

\end{document}